# MITIGATING RADIATION IMPACT ON SUPERCONDUCTING MAGNETS OF THE HIGGS FACTORY MUON COLLIDER[*†]

N.V. Mokhov[#], Y.I. Alexahin, V.V. Kashikhin, S.I. Striganov, I.S. Tropin, A.V. Zlobin

*Fermi National Accelerator Laboratory, Batavia IL 60510-5011, USA*

## Abstract

Low-energy medium-luminosity Muon Collider is being studied as a possible Higgs Factory. Electrons from muon decays will deposit more than 300 kW in superconducting magnets of the Higgs Factory collider ring. Based on the developed detailed MARS15 model and intense simulations, a sophisticated radiation protection system was designed for the entire collider ring to bring the peak power density in the superconducting coils well below the quench limit and reduce the dynamic heat deposition in the cold mass by a factor of 100. The system consists of tight tungsten masks in the magnet interconnect regions and elliptical tungsten liners in magnet aperture optimized for each magnet.

[*]Work supported by Fermi Research Alliance, LLC under contract No. DE-AC02-07CH11359 with the U.S. Department of Energy through the DOE Muon Accelerator Program (MAP).

[†]Presented paper at the 5th International Particle Accelerator Conference, June 15-20, 2014, Dresden, Germany

[#]mokhov@fnal.gov

# MITIGATING RADIATION IMPACT ON SUPERCONDUCTING MAGNETS OF THE HIGGS FACTORY MUON COLLIDER*

N.V. Mokhov#, Y.I. Alexahin, V.V. Kashikhin, S.I. Striganov, I.S. Tropin, A.V. Zlobin,
FNAL, Batavia, IL 60510, USA

## Abstract

Low-energy medium-luminosity Muon Collider is being studied as a possible Higgs Factory. Electrons from muon decays will deposit more than 300 kW in superconducting magnets of the Higgs Factory collider ring. Based on the developed detailed MARS15 model and intense simulations, a sophisticated radiation protection system was designed for the entire collider ring to bring the peak power density in the superconducting coils well below the quench limit and reduce the dynamic heat deposition in the cold mass by a factor of 100. The system consists of tight tungsten masks in the magnet interconnect regions and elliptical tungsten liners in magnet aperture optimized for each magnet.

## INTRODUCTION

Recent discovery of the Higgs boson boosted interest to a low-energy medium-luminosity Muon Collider as a Higgs Factory (HF). A preliminary design of the 125 GeV c.o.m. HF muon Storage Ring (SR) lattice, the Interaction Region (IR) layout and superconducting (SC) magnets along with the first results of heat deposition simulations in SC magnets is described in Ref. [1, 2]. It was shown that the large normalized beam emittance and $\beta_{max}$, and the necessity to protect magnets and detector from showers generated by muon decay products require a very large aperture of SC magnets in both the Interaction Region (IR) and the rest of the ring. A preliminary design of the HF storage ring is based on $Nb_3Sn$ SC magnets with the coil aperture ranging from 50 cm in the interaction region to 16 cm in the arc [2]. The coil cross-sections were chosen based on the operation margin, field quality and quench protection considerations to provide an adequate space for the beam pipe, helium channel and inner absorber (liner).

At the 62.5 GeV muon energy and $2\times10^{12}$ muons per bunch, the electrons from muon decays deposit more than 300 kW in superconducting magnets of the Higgs Factory Interaction Region and Storage Ring. This heat deposition corresponds to unprecedented average dynamic heat load of 1 kW/m around the 300-m long ring or a multi-MW room temperature equivalent if the heat is deposited at helium temperature. This paper presents the results of a thorough optimization of the protection system to substantially reduce radiation loads on the HF magnets and backgrounds.

___________________________________________
*Work supported by Fermi Research Alliance, LLC, under contract No. DE-AC02-07CH11359 with the U.S. Department of Energy through the DOE Muon Accelerator Program (MAP).
#mokhov@fnal.gov

## LATTICE, MAGNETS AND DETECTOR MARS15 MODEL

The MARS15 Monte-Carlo code [3,4] is used to address the key issues related to magnet and detector protection from radiation in HF collider. A detailed 3D model of the entire collider ring including IR, the chromaticity correction (CCS) and matching (MS) sections, arc, and machine-detector interface (MDI) has been built.

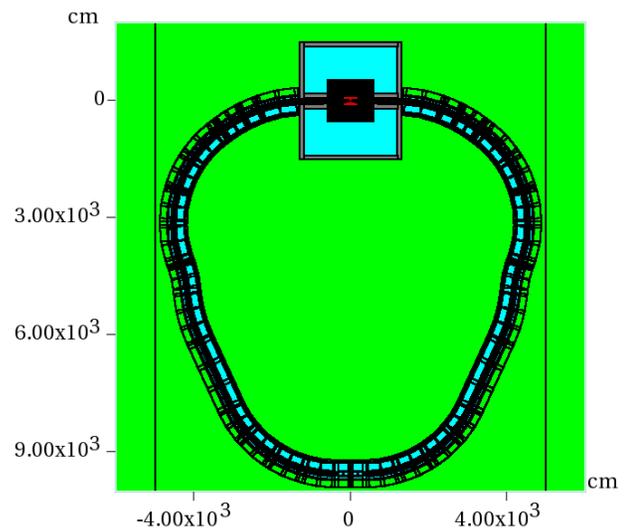

Figure 1: MARS15 model of HF collider including experimental hall and the SiD-like detector at IP.

Fig. 1 shows the MARS model of the 300-m circumference HF with the SiD-like collider detector at the interaction point (IP). Silicon vertex detector and tracker are based on the design proposed for the CMS detector upgrade. The detector geometry description in the GDML format was promptly imported into the MARS15 ROOT geometry model.

The design of all the HF collider ring magnets is based on multi-layer shell-type coils wound using a $Nb_3Sn$ Rutherford cable [2]. The coil apertures were chosen based on the $8\sigma$ beam envelopes and the adequate space needed for the inner absorber (liner), insulating vacuum gap, cold beam pipe, and helium channel. In IR, the first focusing quadrupole Q1, placed at 3.5 m from IP, has the coil aperture of 32 cm. The second and the fourth defocusing quadrupoles with additional dipole coils, the third focusing quadrupole Q3 and first two 8 T dipoles have the coil aperture of 50 cm.

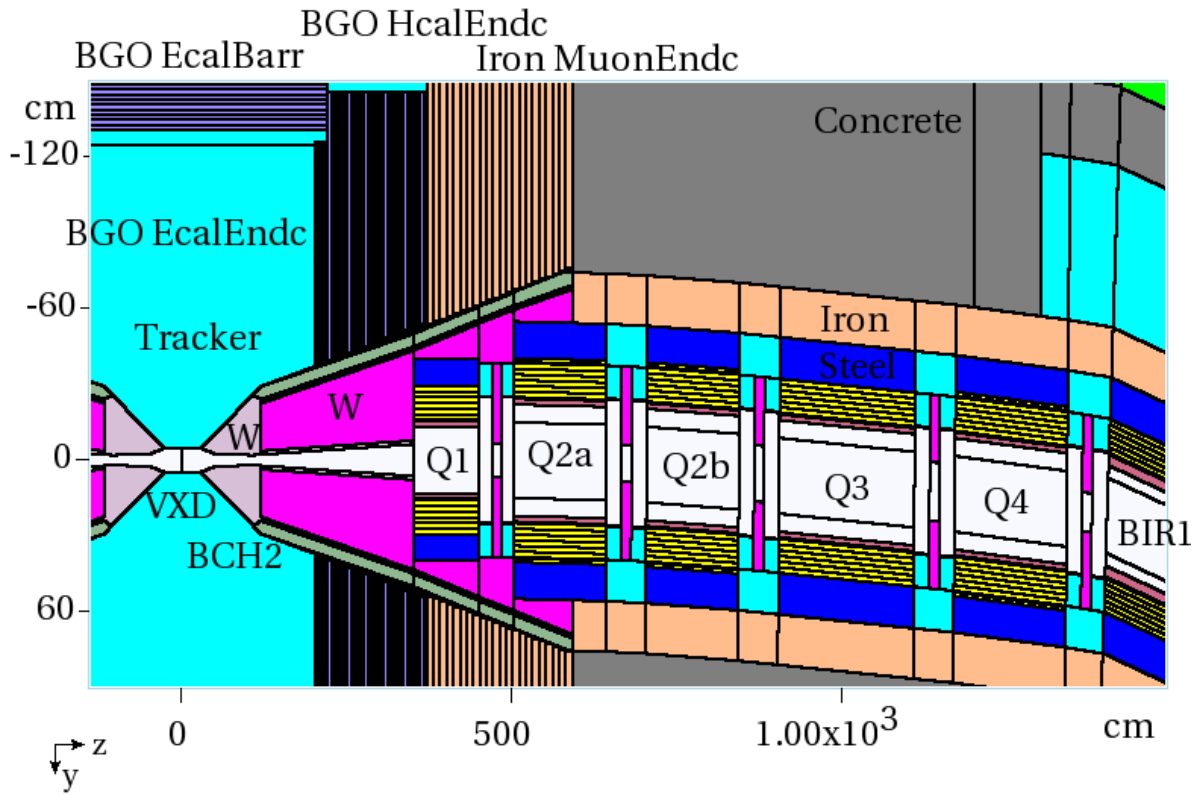

Figure 2: HF MDI MARS15 model with tungsten nozzles on each side of IP, tungsten masks in interconnect regions and tungsten liners inside each magnet.

In the rest of the HF collider ring, the dipoles with $B_{op}$=10 T and the quadrupoles with $G_{op}$=36 T/m and the coil aperture of 27 cm (CCS and MS) and 16 cm (Arc) are used.

The cross-sections of the dipole and quadrupole coils were optimized to achieve the required nominal field or field gradient with sufficient operation margin and good field quality in the beam area. The magnets provide operation margins of 20-50% (in some cases up to 80%). All the magnet details including material properties, were implemented into the MARS15 model along with the fine-meshed magnetic field maps.

## MAGNET PROTECTION SYSTEM

The first studies of radiation heat depositions and doses in the HF collider ring have shown [1, 2] that to provide the operational stability and adequate lifetime of the HF SC magnets the values of radiation load have to be reduced by at least a factor of 100. It was shown in early studies [5, 6], that the practical way to protect SC magnets of a muon collider ring against electromagnetic showers induced by electrons from muon decays is to limit the magnet lengths by 2 to 4 m, install the tight tungsten masks in the magnet interconnect regions, and place thick tungsten liners inside the magnet apertures. The described radiation protection system concept allows reaching the following goals. It (1) provides reduction of the peak power density in the $Nb_3Sn$ cable to ~1.5 mW/g which is below the $Nb_3Sn$ superconductor quench limit with an appropriate safety margin; (2) keeps the lifetime peak dose in the innermost layers of insulation below 20-40 MGy; (3) reduces the average dynamic heat load in the cold mass to the level of ~10 W/m acceptable for cryogenic system; and (4) suppresses the long-range component of the detector background. As a result of massive MARS15 simulations, such a system has been designed to fulfil these constraints and to provide at least a $8\sigma_{x,y}$ beam envelope for muons for up to 2000 turns.

Fig. 2 presents a fragment of the radiation protection system for the IR. The system parameters have been individually optimized for each magnet and interconnect region in the 300-m circumference HF collider ring. As an example, a thickest tungsten liner for one of the hottest dipole magnets in CCS at 24.2 m from IP is shown in Fig. 3. The liner is 4.1-cm thick horizontally and 2-cm thick vertically and is cooled at 60-80 K. It follows radially by the support structure, stainless steel beam pipe, Kapton insulation, liquid helium channel and 4.5 K $Nb_3Sn$ coil.

Calculated isocontours of the power density at the longitudinal maximum of this BCS1 dipole are shown in Fig. 4. The shielding effect of the tungsten mask and liner in the magnet aperture is clearly seen. The peak power density in BCS1 dipole - thanks to the described protection components - is less than 1 mW/g. Similarly, the peak power density in all the SC coils around the ring is reduced to the target value of 1.5 mW/g or less.

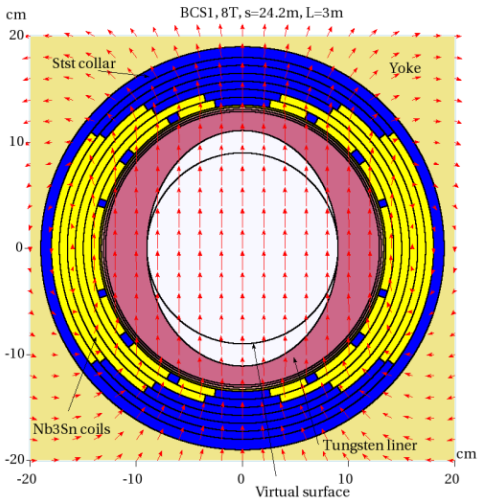

Figure 3: Elliptical tungsten liner inside BCS1 8 T dipole.

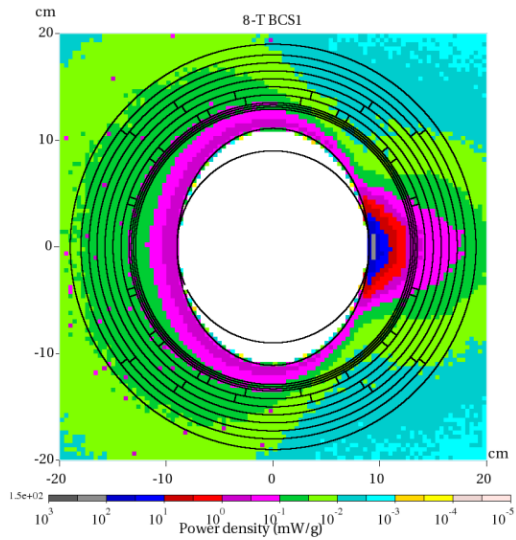

Figure 4: Power density isocontours in BCS1 dipole.

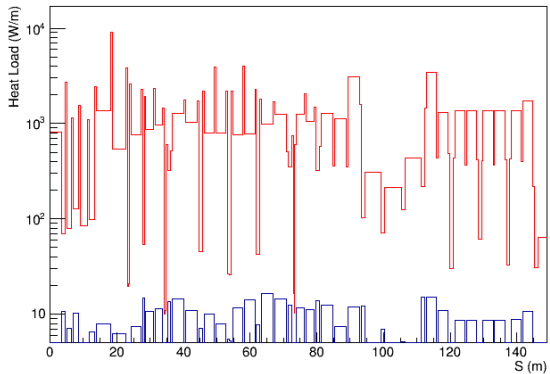

Figure 5: Dynamic heat load in SC magnets along the half of the HF collider for the tungsten masks and liners at 60-80 K (red) and cold mass at 4.5 K (blue).

Dynamic heat loads on the SC magnet cold mass define a capacity of the collider cryoplant and operation costs. An acceptable level of the dynamic heat load is 10 W/m or less at the 4.5 K liquid helium temperature. It means that the magnet protection system needs to reduce the average load to the cold mass 100 times, from the original 1 kW/m. The system described above was designed under this constraint and results are shown in Fig. 5. The average load to the cold mass is below the desirable 10 W/m value. Although in some magnets the heat load level is up to 15 W/m which is higher than average, it is still tolerable for cryogenic system at 4.5 temperature level. The remaining average power dissipation of ~990 W/m is intercepted by the tungsten masks and liners cooled at 60-80 K.

The designed tungsten masks and liners are positioned at the edge of the $8\sigma_{x,y}$ beam envelope or farther. The analysis of related resistive wall impedance and beam stability has shown [7] that one can expect some small (a few percent) growth of an initial perturbation after 1000 turns, so there is a safety factor about one hundred for the transverse plane instabilities. For the longitudinal plane, the magnet protection system can result in up to ~30% of the energy widening. This effect could be probably reduced by means of the second harmonic RF. In any case, thin conducting tapers will be included in the system for a smooth transition between masks and liners.

## CONCLUSION

First time a detailed 3D MARS15 model of the entire HF collider ring including IR, chromaticity correction and matching sections, arc, and machine-detector interface has been built and studied. A sophisticated radiation protection system based on tight tungsten masks in the magnet interconnect regions and optimized elliptical tungsten liners in magnet apertures was designed for the HF collider ring. This system allowed reducing the peak power density in the superconducting coils well below their quench limit and the dynamic heat deposition in the magnet cold masses to the level tolerable by cryogenic system. The obtained results confirm the possibility of radiation protection of $Nb_3Sn$ superconducting magnets in a Higgs Factory muon collider ring.